\documentclass{article}
\usepackage{LaThuileFPSpro}
\usepackage{epsfig,psfrag,graphicx,bm}
\begin{document}
\title{SEARCHES FOR PARTICLE DARK MATTER WITH THE GLAST LARGE AREA TELESCOPE}
\author{
  Jan Conrad        \\
  {\em KTH-Stockholm/High Energy Astrophysics and Cosmology Center (HEAC)}\\ {\em AlbaNova University Centre, 10691 Stockholm} \\
  {Representing the GLAST-LAT collaboration}        \\
 }
\maketitle

\baselineskip=11.6pt

\begin{abstract}
The Large Area Telescope (LAT), one of two instruments on the Gamma-ray Large Area Space Telescope (GLAST) mission, scheduled for launch by NASA in 2007, is an imaging, wide field-of-view, high-energy gamma-ray telescope, covering the approximate energy range from 20 MeV to more than 300 GeV. Annihilation of Weakly Interacting Massive Particles, predicted in many extensions of the Standard Model of Particle Physics, may give rise to a signal in gamma-ray spectra from many cosmic sources. In this contribution we give an overview of the searches for WIMP Dark Matter performed by the GLAST-LAT collaboration.
\end{abstract}
\newpage
\section{Introduction}
There is compelling experimental evidence for a dark component of the matter density of the Universe from observation of on many different scales such as galaxies, galaxy clusters and cosmic background radiation\cite{Bertone:2004pz}. The questions of what constitutes this Dark Matter is one of the great mysteries of modern physics. One of the most promising candidates for the Dark Matter is a Weakly Interacting Massive particle (WIMP). WIMPs can be detected indirectly via their annihilation products, in particular neutrinos, anti-protons, 
positrons and gamma-rays.\\

\noindent
The spectrum of gamma-rays due to WIMP annihilation can be constructed in a ``generic'' fashion, i.e. almost independent of underlying physics model.  The light yield per annihilation is given by:
\begin{equation}
\frac{dN_\gamma}{dE}=\frac{dN_\textrm{cont}}{dE}(E)+\sum_X b_{\gamma X} n_\gamma \delta\left(E - m_\chi(1-m_\chi^2/4m_\chi^2)\right)
\label{eq:yield}
\end{equation}
The first term is the contribution from WIMP annihilations into the full set of tree-level final states, containing fermions, gauge or Higgs bosons, whose fragmentation/decay chain generates photons predominantly via pion decay. For Majorana fermion WIMPs light fermions are suppressed so that the dominant fermionic annihilation products will be $b\bar{b},t\bar{t}$ and $ \bar{\tau}\tau$. The second term is a line originating from annihilation into a two particle final state. As WIMPs are non-relativistic, the photon energy is fixed by the mass of the WIMP, $m_\chi$ and the mass of the other particle X (for example a Z boson), $b_{\gamma X}$ is the branching fraction and $n_\gamma$ is the number of photons per annihilation, i.e. two for the all $\gamma$ final state, one for the others. For the 2 $\gamma$ final state the line is centered on energy $E=M_\chi$. This process is forbidden on tree level with a branching fraction of usually $10^{-3}$ to $10^{-4}$. In addition to the light yield and branching functions the flux depends on the velocity averaged cross-section and the WIMP mass.\\
 
\noindent
In this note we will give a short description of the GLAST project and summarize the searches for Dark Matter envisaged to be performed with GLAST. As examples, we will discuss the potential for GLAST to detect Galactic Dark Matter satellites and a possible signal in the extragalactic background flux.\\

\noindent
For the sensitivities presented in this paper we assume all annihilations to lead to $b\bar{b}$ and 2$\gamma$, the latter with a branching faction of $10^{-3}$. For exclusion of specific models we refer the reader to the contribution by Morselli and Lionetto to the First International GLAST symposium which was held in Stanford, USA, in February 2007\cite{GLASTsymp}.

\section{The Large Area Telescope of GLAST}
GLAST\cite{Michelson:2001yj}, which is part of the NASA's office of Space and Science strategic plan, is an international space mission that will study cosmic $\gamma$-rays in the energy range 20 MeV - 300 GeV.  This mission is realized, as a close collaboration between the astrophysics and particle physics communities, including institutions in the USA, Japan, France, Germany, Italy and Sweden. The main instrument on GLAST is the Large Area Telescope (LAT) complemented by a dedicated instrument for the detection of gamma-ray bursts (the Gamma-ray burst monitor, GBM). The baseline LAT is modular, consisting of a 4 $\times$ 4 array of identical towers. Each 40 $\times$ 40 cm$^2$ tower comprises a tracker, calorimeter and data acquisition module. The tracking detector consists of 18 xy layers of silicon strip detectors. This detector technology has a long and successful history of application in accelerator-based high-energy physics. It is well-matched to the requirements of high detection efficiency ($>$99\%), excellent position resolution ($<$60 $\mu$m), large signal/noise ($>$20), negligible cross-talk, and ease of trigger and readout.  Compared to its predecessor EGRET\cite{EGRET}, the LAT (Large Area Telescope) will have a sensitivity exceeding that of EGRET by at least a factor of 50 (at 100 MeV), the energy range will be extended by a factor 10 and the energy (GLAST: 6 \% at 10 GeV) and angular resolutions (GLAST $PSF_{68\%}< 0.1^{\circ}$ at 10 GeV) are improved by a factor of at least two. The improvement in sensitivity is partly due to the choice of silicon tracking detectors instead of the spark-chambers used in EGRET, which reduces the dead-time by more than three orders of magnitude. GLAST is now integrated in the space-craft and undergoes final testing. The launch of GLAST is scheduled for December 2007.\\

\noindent
The main science targets are (1) to understand the mechanisms of particle acceleration in active
galactic nuclei, pulsars, and supernova remnants (2) to resolve the gamma-ray sky; unidentified sources and diffuse emission (3)
determine the high-energy behavior of gamma-ray bursts and transients, and finally (4) to probe dark matter and early Universe.

\section{LAT searches for Dark Matter}
The GLAST-LAT collaboration pursues complementary searches for Dark Matter each presenting its own challenges and advantages. In table \ref{tab:searches} we summarize the most important ones.\\

\noindent
The center of our own galaxy is a formidable astrophysical target to search for a Dark Matter signal, the reason being that simulations of Dark Matter halos predict high densities at the center of the galaxy and since the WIMP annihilation rate is proportional to the density squared, significant fluxes can be expected. On the other hand, establishing a signal requires identification of the high energy gamma-ray sources which are close (or near) the center\cite{Mayer-Hasselwander:1998hg} and also an adequate modeling of the galactic diffuse emission due to cosmic rays colliding with the interstellar medium. The latter is even more crucial for establishing a WIMP annihilation signal from the galactic halo.\\  

\noindent
Due to the 2$\gamma$ production channel, a feature in the spectrum from the various astrophysical sources would be the gamma-ray line placed at the WIMP mass. This is a ``golden'' signal, in the sense that it would be difficult to explain by an astrophysical process different from WIMP annihilation. Also it would be free of astrophysical uncertainties, since the background can be determined from the data itself. However, since the 2$\gamma$ channel is loop-suppressed, the number of photons will be very low.\\

\noindent
In the following sub-sections, we will give a short description of two of the performed searches: (1) the search for cosmological annihilations of WIMPs and (2) the search for galactic satellites.\\

\begin{table}[t]
  \centering
  \caption{ \it Summary of the different searches for Particle Dark Matter undertaken by the GLAST-LAT collaboration. For reference we include the contributions to the 1$^{st}$ International GLAST Symposium describing the respective analyses\cite{GLASTsymp}}
  \vskip 0.1 in
  \begin{tabular}{|l|l|l|l|} \hline
   Search  &  advantages & challenges & GLAST \\
           &             &            & Symp. \\
    \hline
    \hline
    Galactic     & good            & Difficult source & Morselli\\ 
    center      & statistics       & id, uncertainties  in & et al. \\ 
                &                  & diffuse background & \\ \hline 
    Satellites & Low background,           & low           & Wang  \\
               & good source identification & statistics & et al.   \\   \hline
    Galactic  & Large           & Uncertainties  & Sander  \\ 
    halo       & statistics      & in diffuse     & et al.      \\ 
                &                 & background     &             \\ \hline
    Extra      & Large           & Uncertainties in diffuse & Bergstr\"om   \\ 
    galactic   & statistics      & background,   & et al. \\
               &                 & astrophysical               &  \\ 
               &                 & uncertainties               &  \\ \hline
    Spectral   & No astrophysical & low         & Edmonds \\
    lines      & uncertainties    & statistics  & et al.         \\    
               & ``golden'' signal  &           &               \\
    \hline
  \end{tabular}                  
  \label{tab:searches}
\end{table}

\subsection{Cosmological WIMP annihilation}
Pair annihilation WIMP Dark Matter into high energy photons taking place in dark matter halos at all redshifts might contribute to the extragalactic diffuse gamma-ray radiation. The $\gamma$- annihilation channel would give rise to a distinct feature in the spectrum, a line which is distorted by the integration over all cosmological redshifts. 
The number of photons collected by a detector per unit area-time-energy and solid angle on the sky, originating from WIMP annihilations accumulated over all redshifts, can be calculated by\cite{Ullio:2002pj}:
\begin{equation}
\frac{d\phi_\gamma}{dE_0}=\frac{\sigma v}{8\pi}\frac{c}{H_0}\frac{\bar{\rho}_0^2}{M_\chi^2}\int\, dz(1+z)^3 \frac{\Delta^2(z)}{h(z)}\frac{dN_\gamma(E_0(1+z))}{dE}e^{-\tau(z,\,E_0)}.
\label{eq:flux}
\end{equation}
where particle physics determines the cross section $\sigma$, the WIMP mas $M_\chi$ and the gamma-yield per annihilation given in equation \ref{eq:yield}.  The quantity $\Delta^2(z)$ describes the averaged squared over density in halos, as a function of redshift and $\bar{\rho}_0$ is the present day mean density. The annihilation rate is proportional to the dark matter density squared, which means ``clumpiness'' can significantly enhance the possible signal from WIMP annihilation.\\

\noindent
The extragalactic gamma-ray signal is strongly affected by absorption in the inter-galactic medium, especially at high energies, dominantly by pair production of GeV-TeV photons on infrared/optical background. The absorption is parameterized by the parameter $\tau$, the optical depth. We include the effect of absorption using parameterizations of the optical depth as function of both redshift and observed energy\cite{Primack:2000xp}. More recent  calculation of the optical depth\cite{Stecker:2006eh} do not alter our results significantly. The Hubble parameter enters via the present day value, $H_0$, and the dimensionless quantity $h(z)$, which depends on the energy content of the Universe which changes with redshift. For these values we have used the results from the WMAP three-year data\cite{Spergel:2006hy}.\\

\begin{center}
\begin{figure}[htbp]
\includegraphics[width=7cm,height=7cm]{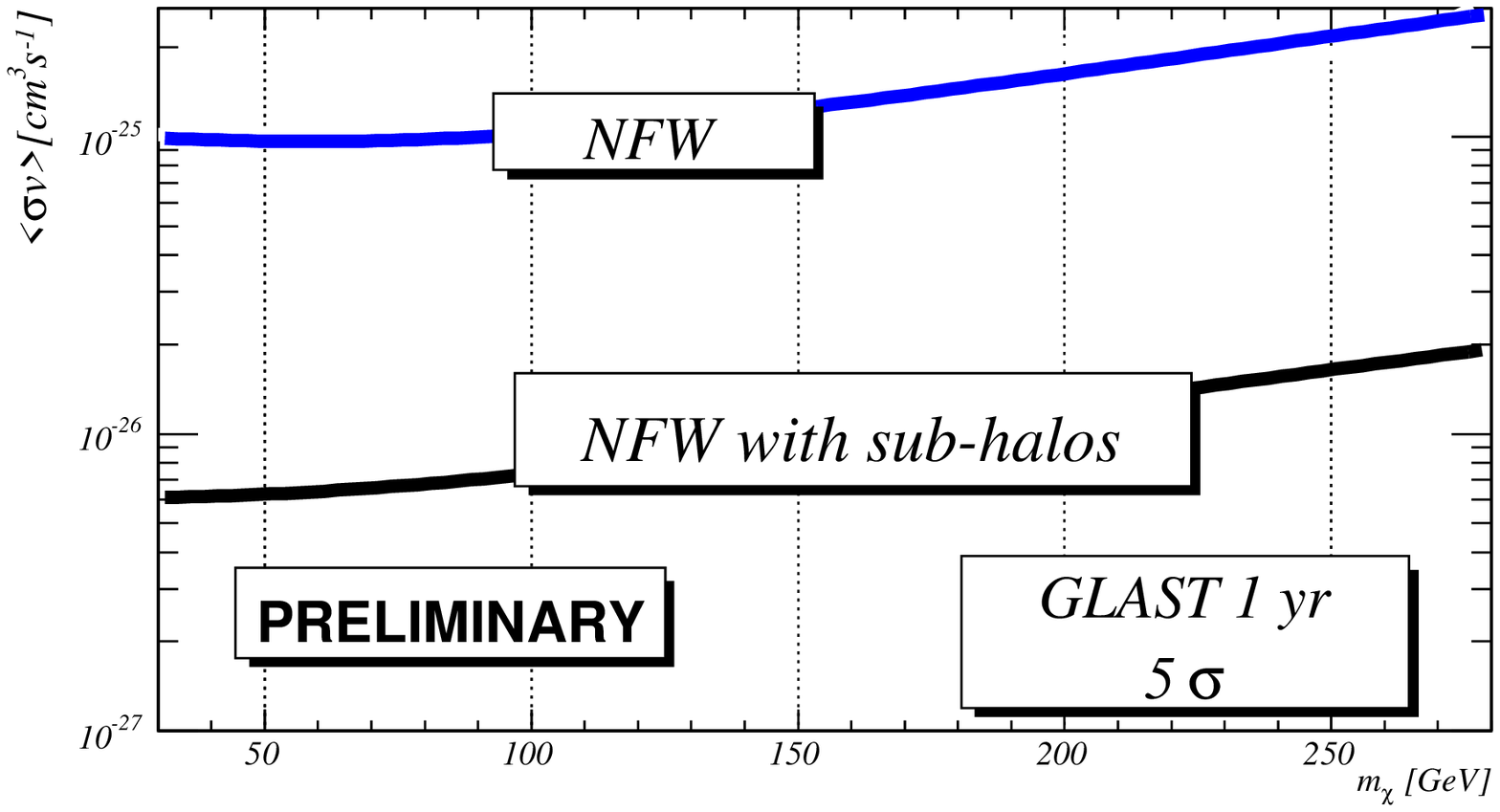}
\includegraphics[width=7cm,height=7cm]{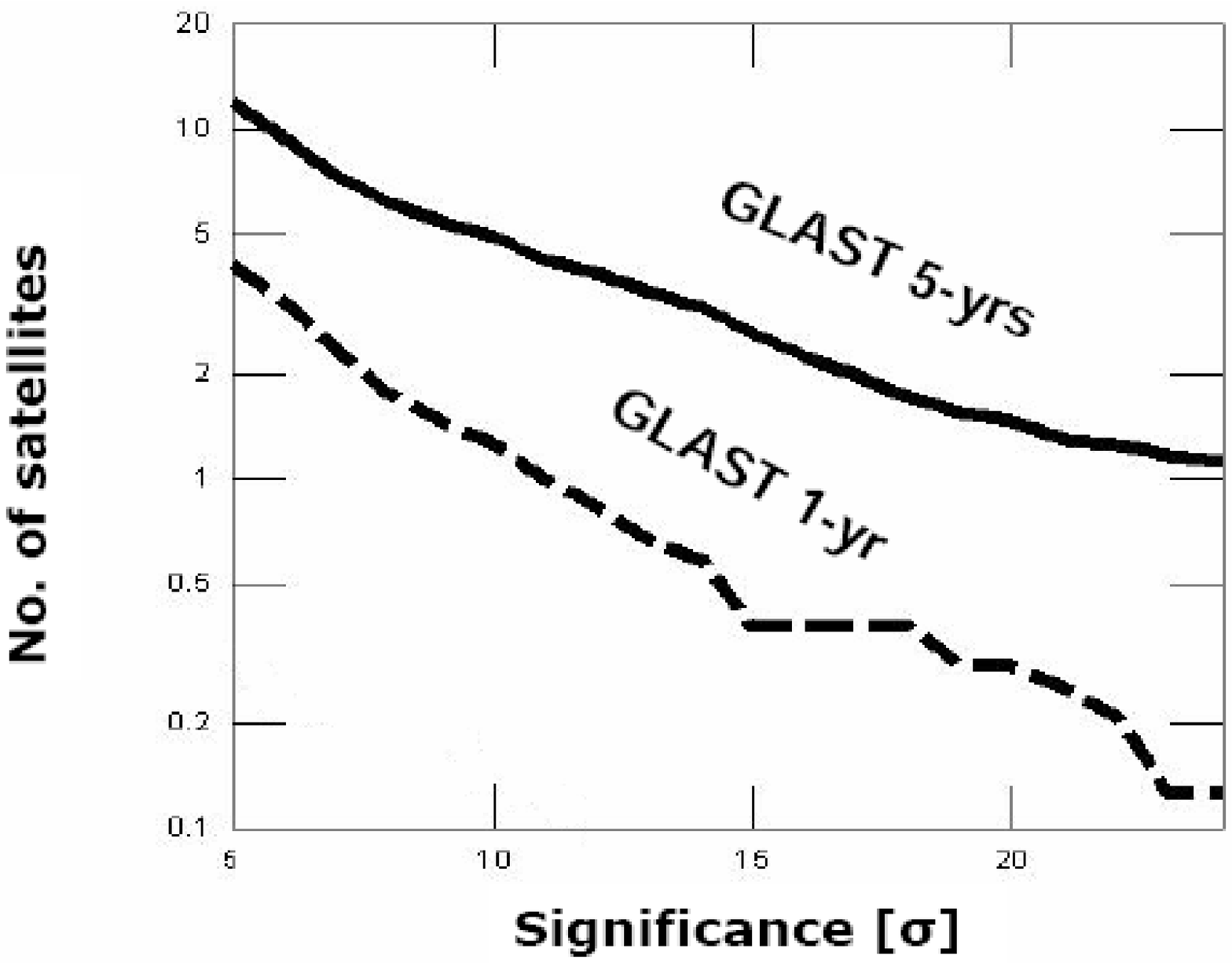}
\caption{{\it Left panel:} 5$\sigma$ exclusion curves for one year of GLAST simulated data. NFW denotes the Navarro Frank White profile, NFW plus sub-halos assumes a substructure in the halos. {\it Right panel:} The number of satellites that could be detected as a function of required significance. See text for more details}
\label{fig:result}
\end{figure}
\end{center}

\noindent
To obtain preliminary estimates of the GLAST senstivity to this type of signal, fast detector simulations were performed for a generic model of WIMPs annihilating into 2$\gamma$ and into $b\bar{b}$ as described in the previous section. A $\chi^2$ analysis was performed, assuming that the background consists of unresolved blazars\cite{Ullio:2002pj} to obtain a sensitivity plot in $<\sigma v>$ vs $M_\chi$. The WIMP signal was computed using the Navarro-Frank-White (NFW) profile\cite{Navarro:1996gj} for normalization. According to N-body simulations, within larger halos there might exist smaller, bound halos that have survived tidal stripping. Although not as numerous as the primary  halos the substructure halos arise in higher density environments which makes them denser than their parent halo. In addition to a smooth NFW profile, we therefore also consider the case of NFW profile with subhalos. Here, we assumed that they constitute $5\%$ of the mass and have three times the concentration parameter of the parent halo.  The distribution of concentration parameters, as a function of halo mass are inferred from N-body simulations\cite{Bullock:1999he}. The result (see figure \ref{fig:result}, right panel) shows that GLAST should be sensitive to total annihilation cross-sections of the order $10^{-26}-10^{-25}$ cm$^3$ s$^{-1}$, depending on the halo model. One should note that this estimate neglects contributions of instrumental background, uncertainties introduced by the analysis (where point-sources and galactic diffuse emission have to be taken into account) and finally that the extragalactic background spectrum from astrophysical sources is very uncertain, especially at high energies. 

\noindent

\subsection{Searches for Dark Matter satellites}
Subhalos which, as seen in the previous section, lead to a signficant enhancement of the WIMP annihilation induced flux in the extragalactic background would be present also in our own Galaxy. They could be detectable as Dark Matter satellites. We used simulations of Dark Matter satellite formation\cite{Taylor:2004gq} combined with a fast detector simulation to predict the number of satellites detectable by GLAST. The DM satellite distribution is roughly spherically symmetric about the galactic center, with most of the observable satellites located at high galactic latitude (i.e. relatively low background). For this particular study we used reference\cite{Cillis:2005bd} to estimate the background as the point source subtracted sky map above 1 GeV \footnote{This is probably a conservative assumption, since GLAST will be able to resolve many more point sources.}. The significance of detection was then calculated by estimating the number of signal events within the satellite tidal radius (or the PSF 68 \% containment radius, whichever was bigger) divided by the square root of background events within the same radius. Figure \ref{fig:result}, right panel,  shows the number of satellites which could be detected by GLAST above a given signficance in 1 and 5 years of GLAST operation. Here a 100 GeV WIMP with a velocity averaged cross-section $<\sigma v> = 2.3 \cdot 10^{-26}$ cm$^3$ s$^{-1}$ was assumed\footnote{a line contribution was neglected due to negligible statistics}. Under this assumptions GLAST will be able to detect a few highly significant satellites during 5 years operation. It should be noted that the true significance of detection will also have to take into account the fact that Dark Matter satellites need to be distinguished from other astrophysical sources,the most difficult probably being pulsars (see\cite{Baltz:2006sv} for further discussion). 

\section{Conclusions}
In this note we summarize the searches for particle Dark Matter to be performed with the GLAST-LAT instrument. Several complementary astrophysical sources will be examined, each presenting its own advantages and challenges. Those mentioned here are the galactic center, galactic halo, extragalactic background and galactic satellites. The golden signal for presence of particle dark matter would be a gamma-ray line at the mass of the WIMP. We showed that the GLAST-LAT has the possibility to detect a contribution of WIMP annihilation at all redshift to the extragalactic gamma-ray background. We also showed that galactic Dark Matter satellites can potentially be detected by the GLAST-LAT. GLAST is now integrated on the space-craft and undergoing final testing. The launch is foreseen for December 
2007.

\section{Acknowledgments}
I would like to thank all members of the Dark Matter and New Physics Working Group of the GLAST-LAT who contributed to this note.

\end{document}